\documentclass[pre,aps,floats,superscriptaddress,floatfix,twocolumn]{revtex4-1}
\usepackage{amssymb,amsmath}
\usepackage[normalem]{ulem}
\usepackage{amsmath,amssymb}
\usepackage{graphicx}
\usepackage{psfrag}
\usepackage{mathtools}
\usepackage{color}
\usepackage{stfloats}

\def\beq{\begin{equation}}
\def\eeq{\end{equation}}
\def\bea{\begin{eqnarray}}
\def\eea{\end{eqnarray}}

\begin{document}
\title{Nonequilibrium steady states in a closed inhomogeneous
asymmetric exclusion process with particle nonconservation  }
\author{Bijoy Daga}\email{bijoy.daga@saha.ac.in}
\author{Souvik Mondal}\email{souvik.mondal@saha.ac.in}
\affiliation{Condensed Matter Physics Division, Saha Institute of
Nuclear Physics, Calcutta 700064, India}
\author{Anjan Kumar Chandra}\affiliation{Department of Physics, Malda College, 
Malda, West Bengal, Pin: 732101, India}\email{anjanphys@gmail.com}
\author{Tirthankar Banerjee}\email{tirthankar.banerjee@saha.ac.in}
\author{Abhik Basu}\email{abhik.basu@saha.ac.in,abhik.123@gmail.com}
\affiliation{Condensed Matter Physics Division, Saha Institute of
Nuclear Physics, Calcutta 700064, India}

\date{\today}

\begin{abstract}
We study asymmetric exclusion processes (TASEP) on a nonuniform
one-dimensional ring  consisting of two segments having unequal 
hopping rates, or {\em defects}. We allow weak particle nonconservation via 
Langmuir kinetics (LK), that are
parameterised by generic unequal attachment and detachment rates. For an 
extended defect, in the thermodynamic limit the system generically displays 
inhomogeneous density profiles in the steady state - the faster segment is 
either in a phase with spatially varying density having no density 
discontinuity, or a phase with a discontinuous density 
changes. Nonequilibrium phase transitions 
between them are controlled by the inhomogeneity and LK. The slower segment  
displays only macroscopically uniform bulk density 
profiles in the steady states, reminiscent of the maximal current phase of 
TASEP but with a bulk density generally different from half.  With a point 
defect, there are low and 
high density spatially uniform phases as well, in addition to the inhomogeneous 
density profiles observed for an extended defect. In all the cases, it is 
argued that the the mean particle density in the steady state is controlled 
only by the ratio of the LK attachment and detachment rates. 
\end{abstract}

\maketitle

\section{Introduction}
The effects of nonuniformities or disorders on the macroscopic properties of 
equilibrium 
systems are well understood by now within the general framework of statistical 
mechanics~\cite{equi-disord}. In contrast, the effects of quenched disorders on 
the dynamics and nonequilibrium steady states (NESS) of out of equilibrium 
systems are much less understood~\cite{noneq-quench}. 
Totally asymmetric simple exclusion process (TASEP) and its variants with open 
boundaries in one dimension (1D) serve as simple models of restricted 1D 
transport~\cite{review}. Natural realisations of TASEP include 
motions in nuclear pore complex of
cells~\cite{nuclearpore}, motion of molecular motors along 
microtubules~\cite{molmot}, fluid flow in artificial
crystalline zeolites~\cite{zeo} and protein synthesis by messenger
RNA (mRNA) ribosome complex in cells~\cite{albertbook}; see, e.g., 
Refs.~\cite{review} for basic reviews
on asymmetric exclusion processes. Subsequently, studies on TASEP with particle nonconservation in the 
form of on-off Langmuir kinetics (LK)~\cite{erwin-lk}
in the limit, when LK competes with the hopping movement of TASEP reveals unusual phase 
coexistences in the NESS, not found in pure TASEP. Furthermore, 
open TASEPs with defects~\cite{tasep-def}, both point and extended,
have been studied, which investigated the 
effects of the defects on the NESS and currents.
Open TASEP with a single point defect along with LK is shown to 
display a variety of phases and phase coexistences as a result of the 
competition between the defect and LK~\cite{erwin2}.
\par Studies of TASEP in a closed ring are relatively few and far between. Translational 
invariance ensures that TASEP in a homogeneous ring, with or 
without LK, produces a homogeneous density profile in the steady state. 
Nonuniform or inhomogeneous steady states 
are expected only with explicit breakdown of the translation invariance, e.g., 
by means of quench disorder
in the hopping rates at different sites. Exclusion processes in closed 
inhomogeneous rings with strict particle number conservation have been shown to 
display inhomogeneous NESS~\cite{Mustansir1,niladri,tirtha1}. More recently, TASEP in a ring 
with quench disordered hopping rates together with nonconserving LK having equal rates 
of particle attachment and detachment 
has been studied in Ref.~\cite{LKTASEP_K=1}. Very unexpectedly, the model 
admits only macroscopically inhomogeneous NESS, in the forms of two- and 
three-phase coexistences, regardless of extended or point defects, a feature 
that has been explained in general terms in Ref.~\cite{LKTASEP_K=1}. Equal 
attachment and detachment rates, as used in Ref.~\cite{LKTASEP_K=1}, are actually 
an idealisation and simplification. In typical physical realisations of this model, e.g., 
vehicular/pedestrial traffic and ribosome translocations along closed 
mRNA loops in the presence of defects with 
particle nonconservation, attachment and detachment rates are generically 
expected to be unequal. Thus, it is important to find how unequal rates will 
affect the results of Ref.~\cite{LKTASEP_K=1}, or, the robustness of the 
results in Ref.~\cite{LKTASEP_K=1} against variation in the ratio $K$
of the attachment and detachment rates. In this work, we set out to study this 
question by considering TASEP in a two-segment ring having unequal hopping 
rates with LK. How LK dynamics with $K\neq 1$ affects the phases of an 
open TASEP are studied in Ref.~\cite{erwin-lk}. Here, we analyse how the 
interplay of LK with $K\neq 1$ and quenched inhomogeneity affects the NESS of a 
TASEP on a ring. The latter admits only strictly uniform density due to the 
particle number conservation and translational invariance, and has no phase 
transitions. In the present study, we find that with an extended defect, in 
the NESS the faster segment displays either an inhomogeneous (spatially nonuniform) 
phase with {\em no discontinuity} in the density 
(hereafter {\em continuous density phase}, or CDP) and a phase with a 
discontinuous change or a 
shock (hereafter {\em shock phase}, or SP) in the density profile  for all 
values of the defect 
strength (i.e., the hopping rate along the extended defect) and the ratio of 
the attachment and detachment rates. Steady state density profiles in the CDP 
are found to have three parts - two parts with nonzero spatial gradients of the 
density and an intervening part having a constant magnitude of 
$n_0=K/(1+K)$, i.e., with zero slope. In general, $n_0\neq 1/2$ for $K\neq 1$, 
unlike the corresponding result in Ref.~\cite{LKTASEP_K=1}. The CDP and SP, respectively, are 
the direct generalisations of the three- and two- phase coexistences for $K=1$~\cite{LKTASEP_K=1}. By tuning 
the model parameters, 
nonequilibrium transitions between CDP and SP are observed. The steady 
state density in the slower 
segment is {\em always} macroscopically uniform. Being controlled solely by the LK attachment and detachment 
rates, the value of the steady state density in the slower segment is $n_0$. This is reminiscent of 
the maximal current (MC) phase of TASEP, but with a current generically less than 1/4 for $K\neq 
1$. In contrast, for a point defect, the model can 
display spatially macroscopically uniform  steady states for either $K$ very 
small or very large, yielding either 
low density (LD) or high density (HD) phases. For intermediate values of $K$, 
the model exhibits nonuniform phases - either CDP or SP, as with an 
extended defect. Nonequilibrium 
transitions between all the three types of phases are again controlled by the 
LK attachment-detachment rates and inhomogeneity. Lastly, in all the 
phases of both extended and point defects, the 
mean particle density in the NESS of the system is argued to be $n_0$, i.e., 
controlled only by $K$. Our study is a close-ring analogue of the model  in 
Ref.~\cite{erwin-lk} that considered an open system. Our results, as described 
in details below, manifestly reveals the the significance of the ring geometry 
in controlling the NESS.
The rest of the article is organised as follows: In Sec.~\ref{the_model}, we define 
our model. Then, a short review of the results for $K=1$ is presented in 
Sec.~\ref{rev}. In Secs.~\ref{1a} and \ref{1b} we analyse the NESS of the model 
with an extended and a point defect, respectively. We summarise in 
Sec.~\ref{summ} at the end.
\section{The Model}
\label{the_model}
\begin{figure}[t!]
\centering \includegraphics[height=7cm]{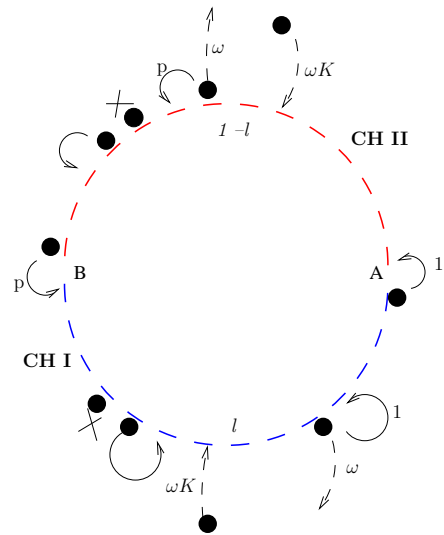}
\caption{(Colour online) TASEP and LK dynamics on a 1D ring with unequal 
hopping rates in CHI 
and CHII; the LK detachment and attachment rates are $\omega$ and $\omega K$, 
respectively; see text.}
\label{model}
\end{figure}

Consider a TASEP of $N$ sites together with LK on a 1D periodic ring with 
spatial inhomogeneity. Owing to exclusion, 
the particle occupation, $n_i$ at  each site $i$ is either $0$ or $1$. In  
TASEP, a particle 
can only hop to its immediate vacant neighbour in one direction, say 
anticlockwise (see Fig.~\ref{model}).  Spatial inhomogeneity
in the model is then introduced by  unequal hopping rates of the 
particles in the two segments, one faster and the other slower, marked CHI 
and CHII respectively in Fig.~\ref{model}. The hopping rates in CHI and CHII, 
respectively, are 1 and $p(<1)$.  Let  
$N_1$ and $N_2$ be the respective 
sizes of CHI and CHII, such that $N_1+N_2=N$. We define $l=N_1/N$ as the 
fraction of the total sites in CHI; $N_1=Nl,\,N_2=(1-l)N$. Note 
that since the hopping rates  do not evolve in time,
the spatial inhomogeneities considered here are quenched. Further, due to LK, a particle can 
detach from a given site with rate $\omega$ or attach to a given site with rate 
$\omega K$. We work in the regime where TASEP competes with 
LK-dynamics, such that  the net flux due to 
TASEP must be comparable to the total flux due to LK. To ensure this, we 
apply scaling relations: $\Omega=\omega N$, where $N$ is a large but finite number and 
study the system in the limit 
$\Omega$ $\sim O(1)$ \cite{erwin-lk, LKTASEP_K=1}. For convenience, we 
label the sites by
a continuous variable $x$ in the thermodynamic limit, defined by $x = i/N , 0 < x 
< 1$. In terms of the rescaled
coordinate $x$, the lengths of CHI and CHII are $l$ and $1-l$,
respectively. The coordinate $x$ rises from junction B ($x=0$) to junction A 
($x=l$) in the anticlockwise direction.
\subsection{Short review of the results for $K=1$}\label{rev}
We briefly review the results for $K=1$~\cite{LKTASEP_K=1}, i.e., with equal 
attachment and detachment rates. The mean-field (MF)
equations  are set up in both 
CHI and CHII. The steady state densities are then calculated by observing 
that current is conserved ${\it locally}$, and hence
current conservation can be applied very close to the junctions $A$ and $B$:
\beq
 n_{I}(y)(1-n_{I}(y))= p n_{II}(y)(1-n_{II}(y)),
 \label{current_cons}
\eeq
where $y=$ junctions $A$ or $B$. The steady state densities  are given by
\beq
 n_{IB}(x) = \Omega x + \frac{1- \sqrt{1-p}}{2}, 
 \label{K=1eq1}
\eeq
\beq
 n_{IA}(x) = \Omega (x-l) + \frac{1 + \sqrt{1-p}}{2}, 
 \label{K=1eq2}
\eeq
such that both Eqs.~(\ref{K=1eq1}) and (\ref{K=1eq2}) depend upon the density  
values at the junctions B and A, respectively. In contrast,
\beq
 n_{I b}(x) = \frac{1}{2}, 
 \label{K=1eq3}
\eeq
and
\beq
n_{II}(x)= \frac{1}{2} 
\label{K=1eq4}
\eeq
are the {\em bulk} solutions independent of the densities at the junctions, and 
are analogue of the MC
phase in an open TASEP. In CHI, let $x_{\alpha}$ and $x_{\beta}$ respectively be the points where the 
left solution $n_{IB}(x)$ and the 
right solution $ n_{IA}(x)$ meets with the bulk solution $n_{Ib}$. It was shown that
\beq
x_{\beta}- x_{\alpha}= l- \frac{\sqrt{1-p}}{\Omega}.
\label{xx_diff}
\eeq
Thus if $x_{\alpha}<x_{\beta}$, the system is in a three phase (LD-MC-HD) 
coexistence.  But if $x_{\alpha}>x_{\beta}$, the system
can be only in a two phase(LD-HD) coexistence. The critical line  separating 
the two phases can be thus determined by setting 
$x_{\alpha}=x_{\beta}$. Further, the global average density of the system is 
always $n_0=1/2$, irrespective of the chosen values 
of $p$ and $\Omega$.
\section{Steady state densities}
We perform  MF analysis of our model, supplemented by  its 
extensive Monte Carlo Simulations (MCS). We separately present our results for 
an extended and a point defect below. We consider only $K<1$; the results 
for $K>1$ may be obtained from those for $K<1$ together with the particle-hole 
symmetry. 
In our MCS studies, we use a 
random-sequential update scheme. We measure the average site density, $\langle 
n_i\rangle$
over approximately $2 \times 10^9$ Monte-Carlo steps  after relaxing the system 
for $10^9$ Monte-Carlo steps. We separately study an extended and a point 
defect. For an extended defect here, 
$0<l<1$, where as for a point defect, $l\rightarrow 1$. In 
Ref.~\cite{erwin-lk}, the authors solved the steady state densities in the 
analogous open system in terms of the Lambert $W$ functions. Here, we use the 
implicit solutions of the densities in the steady state that suffice for our 
purposes.
\subsection{MF analysis and MCS results for an extended defect}\label{1a}
We set up our MF analysis by closely following the logic outlined in 
Ref.~\cite{LKTASEP_K=1}. In 
our MF analysis, we describe the model as a combination of two TASEPs - CHI and 
CHII,  joined with each other at the junctions A and B; see Fig.~\ref{model}. 
Thus, junctions B and A 
are {\em effective}
entry (exit) and exit (entry) ends of CHI (CHII). This allows us to analyse the
phases of the system in terms of the known phases of the open boundary 
LK-TASEP~\cite{erwin-lk}. Without LK, the steady state densities of a TASEP in an inhomogeneous ring are completely 
determined by  the total particle number (a conserved quantity)  in the system 
in the steady state and the inhomogeneity configurations~\cite{Mustansir1,niladri,tirtha1}. 
Due to the nonconserving LK dynamics, however, the particle current here is 
conserved only {\em locally},
since the probability of attachment or
detachment at a particular site vanishes as $1/N$~\cite{erwin-lk,LKTASEP_K=1}. 
Hence, the total particle number in the NESS is not a conserved quantity.
The steady state densities $n_I(x)$ and $n_{II} (x)$ in CHI and 
CHII respectively follow~\cite{erwin-lk,LKTASEP_K=1}
{\small
\beq
 \left(2 n_I(x)-1\right)\frac{\partial n_{I}(x)}{\partial x}-\Omega (1+ K) 
 \left(n_{I}(x)-\frac{K}{1+K}\right) = 0, 
 \label{ch1_den}
\eeq
}
and 
{\small
\beq
p\left(2 n_{II}(x)-1\right)\frac{\partial n_{II}(x)}{\partial x}-\Omega 
(1+K)\left(n_{II}(x)-\frac{K}{1+K}\right)=0.
\label{ch2_den}
\eeq
}
Before we attempt to solve Eqs.~\eqref{ch1_den} and \eqref{ch2_den},  
we define an average density 
$n_0$ in a given NESS that remains a constant on average, although the model dynamics does not admit 
any conservation law for the total 
particle number. In the homogeneous limit of the model, i.e., 
with $p=1$, CHI and CHII are identical and the steady state density is 
spatially uniform, due to the translational invariance for $p=1$ for all 
$K$. 
Equation~\eqref{ch1_den} or \eqref{ch2_den} then yield
\begin{equation}
 n_0=\frac{K}{K+1}.\label{global_den_eq}
\end{equation}
Thus, the average hole density is $1-n_0$ for $p=1$. A global deviation 
of the mean density from $n_0$ 
should indicate either more particles or
more holes entering into the system in the steady state, than that is given
by $n_0$. However, even when inhomogeneity is introduced ($p<1$),
there is nothing that favours either particles or holes, since the 
inhomogeneity that acts as an inhibitor
for the particle current, equally acts as an inhibitor for the hole current.  
Hence, it is not expected to affect the value of $n_0$. This is a major 
result of this work that is in agreement with the MCS studies; see below. 
Notice that this argument does not
preclude any local excess of particles or holes, since the particles and the 
holes move in the opposite
directions, and hence the presence of a defect should lead to excess particles 
on one side and excess holes
(equivalently deficit particles) on the other. This holds for any 
$\Omega$ and  $K$. When $K=1$, $n_0=1/2$, in agreement 
with Ref.~\cite{LKTASEP_K=1}.
Eqs.~\eqref{ch1_den} and \eqref{ch2_den} are first order differential 
equations, each having one constant of integration in their solutions. These may be 
determined by considering the current conservation or ``boundary 
conditions'' at the junctions A and B. In addition, Eqs.~\eqref{ch1_den} and 
\eqref{ch2_den} admit a third spatially constant solution independent of the 
boundaries, given by $K/(1+K)$. Since
CHI has a higher hopping rate ($1>p$), on physical grounds there cannot be pile 
up of particles in CHII behind junction B. 
In the same way, we do not expect an $x$-dependent $n_{II}(x)$ that
decreases  with $x$  from junctions A to B.
Thus, $n_{II}(x)$ should be macroscopically uniform in the bulk. We are then 
left with only one solution $n_{II}(x)=K/(1+K)$. This solution is 
independent of the boundaries A and B, and is thus akin to the maximal current 
(MC) phase of TASEP. There is however a crucial difference: in an MC phase of a 
TASEP, the steady state bulk density reaches its maximum value of 1/2. 
However, with $n_{II}=K/(1+K)$, the
steady state bulk density in CHII is {\em always} less than $1/2$ (with $K < 
1$). 
\begin{figure}[t!]
\centering \includegraphics[height=5.5cm]{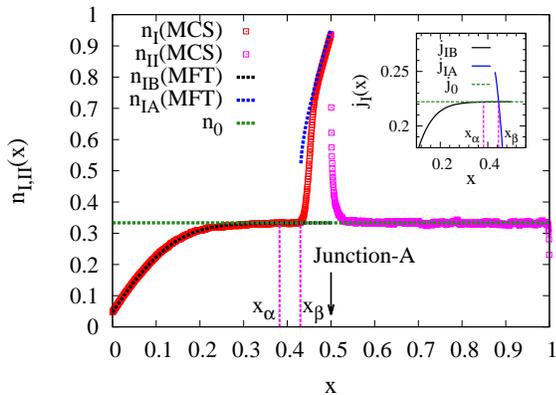}
\caption{(Colour online) Plots of the density profiles in CDP  
($K=0.5,\,p=0.2,\,\Omega=4.5$, $l=1/2,\,N=2000$).  Mean-field solutions 
$n_{IB}(x)$ (black dotted lines), $n_{IA}(x)$ (blue dotted lines) and the corresponding 
MCS results $n_I(x)$ (red squares) are shown. MCS results for $n_{II}(x)$ 
are shown with magenta squares; $x_\alpha$ and $x_\beta$ 
are extracted from MCS data (see text). $n_0$ (green dotted lines) is the 
average density of the system. Inset: Mean-field values  of the 
currents $j_{IB}(x)$, $j_{IA}(x)$ and $j_0(x)$ are plotted; $x_\alpha$ and 
$x_\beta$ are extracted (see text), which match well with their corresponding 
MCS results.}
\label{K0.5_p0.2_om4.5}
\end{figure}
\begin{figure}[t!]
\centering \includegraphics[height=5.5cm]{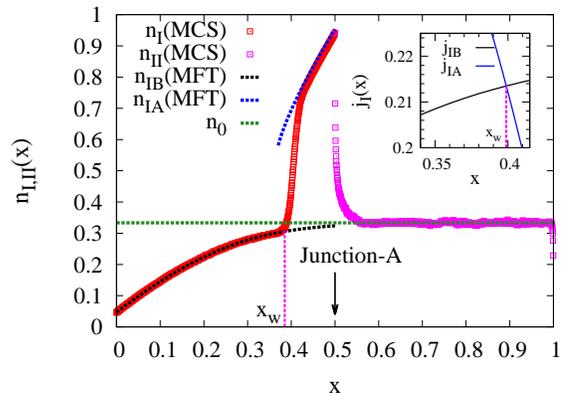}
\caption{ (Colour online) Plots of the density profiles in  SP 
($K=0.5,~p=0.2,~\Omega=2.25,~l=1/2,\,N=2000$).  Mean-field solutions 
$n_{IB}(x)$ 
(black dotted lines), $n_{IA}(x)$ (blue dotted lines) and the corresponding 
MCS results $n_I(x)$ (red squares) are shown. MCS results for $n_{II}(x)$ are shown with magenta squares; $x_w$ is 
extracted from MCS data (see text). $n_0$ (green dotted lines) is the 
average density of the system. Inset: Mean-field values  of the 
currents $j_{IB}(x)$, $j_{IA}(x)$  are plotted; $x_w$ has been extracted (see text), which match well with their corresponding 
MCS results.}
\label{ext_2ph}
\end{figure}
We therefore call it the {\em generalised MC} (GMC) phase~\cite{gmc-k}; see 
also Ref.~\cite{erwin-lk} for analogous results in a similar open system. Now 
consider CHI: since the bulk steady state density in CHII is $K/(1+K)=n_0$,
the average steady state density in CHI must also be $K/(1+K)$, in order to 
have $n_0$ as the mean density in the whole system. Notice that a uniform 
$n_I(x)=n_0$ does solve (\ref{ch1_den}) above.  However, this solution is not 
admissible, as it manifestly violates current conservations at A 
and B. Thus, CHI can only admit macroscopically nonuniform density profiles in 
its NESS. If there are spatially varying LD phases in CHI (monotonically rising 
from B to A, remaining less than 1/2 everywhere), the current conservation at either the 
junctions A and B will be violated. This rules out a spatially varying LD 
phase in CHI. Clearly then, an analogous HD phase in CHI is also ruled out. 
Therefore, on physical grounds one part of the solution for $n_I(x)$ should be 
$<1/2$ (near junction B) and another part $>1/2$ (near A) to maintain current 
conservation at both A and B; see Eqs.~(\ref{left_boundary}) and (\ref{right_boundary}) below. This leaves us with two 
possibilities - (a) the two parts of 
the solutions matching smoothly with the bulk solution $n_I(x)=n_0$, {\em without} any density 
discontinuity (CDP solution), or (b) the two parts {\em do not} meet with the bulk solution, leading 
to a density discontinuity (SP solution), as discussed below. These arguments 
are used below to analyse the phases in the model.
\subsubsection{Phase diagrams and density profiles with an extended defect}
As pointed out in the previous section, CHII is always in the GMC-phase, 
with $n_{II}(x)=n_0$ in the bulk.
Since a pure LD (hence an HD) phase is ruled out in CHI, it should only 
have macroscopically inhomogeneous densities. This means there are no boundary 
layers in CHI at the 
junctions A and B. Then, applying the current 
conservation as given by Eq.~\eqref{current_cons} we get,
{\small
\beq
n_{I}(0)=\alpha_1=\frac{1}{2}\left[1 - \frac{1}{1+K} \sqrt{(1+K)^2 - 4p 
K}\right]<\frac{K}{1+K}, 
 \label{left_boundary}
\eeq
} 
and 
{\small
\beq
n_{I}(l) = \alpha_2= \frac{1}{2}\left[1 + \frac{1}{1+K} \sqrt{(1+K)^2 - 4p 
K}\right]>\frac{1}{2}.
 \label{right_boundary}
\eeq
}
We also obtain the 
general solution to Eq.~\eqref{ch1_den}, 
\beq
\frac{1}{a} \left[ 2 n_I(x)- \left( 1+ \frac{2 b}{a}\right) \ln |a n_I(x)+ b| 
\right]= x +C, 
\label{eq:solution}
\eeq
where $a=\Omega(1+K)$ and $b=-\Omega K$. The constant of integration, $C$ is to 
be determined 
appropriately by using either the boundary conditions \eqref{left_boundary} or 
\eqref{right_boundary}, yielding generally two different solutions. 
Defining $n_{IB}(x)$ and $n_{IA}(x)$ as the 
solutions of Eq. \eqref{ch1_den} corresponding to the boundary conditions 
(\eqref{left_boundary}) and (\eqref{right_boundary}), respectively, we find
\beq
2 \left(n_{IB}(x)-\alpha_1\right)- \left( 1+ \frac{2 b}{a}\right) \ln 
\left|\frac{a n_{IB}(x)+ b}{a \alpha_1+ b}\right|= ax, \label{sol1n1}
\eeq
and 
\beq
2 \left(n_{IA}(x)-\alpha_2\right)- \left( 1+ \frac{2 b}{a}\right) \ln 
\left|\frac{a n_{IA}(x)+ b}{a \alpha_2+ b}\right|= a(x-l).\label{sol2n1}
\eeq
\begin{figure}[t!]
\centering \includegraphics[height=5.5cm]{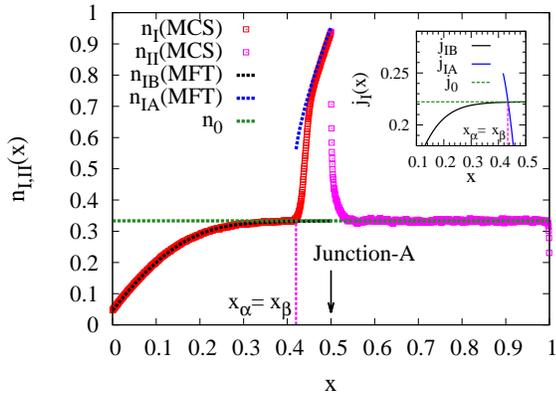}
\caption{ (Colour online) Plots of the density profiles in the borderline case
($K=0.5,\,p=0.2,\,\Omega=3.75,~l=1/2,\,N=2000$).  Mean-field solutions 
$n_{IB}(x)$ 
(black dotted lines), $n_{IA}(x)$ (blue dotted lines) and the corresponding 
MCS results $n_I(x)$ (red squares) are shown. MCS results for $n_{II}(x)$ are shown with
magenta squares; $x_\alpha$ and $x_\beta$ 
are extracted from MCS data (see text). $n_0$ (green dotted lines) is the 
average density of the system. Inset: Mean-field values  of the 
currents $j_{IB}(x)$, $j_{IA}(x)$ and $j_0(x)$ are plotted; $x_\alpha$ and 
$x_\beta$ are extracted (see text), which match well with their corresponding 
MCS results.} 
\label{ext_bound}
\end{figure}
\begin{figure}[t!]
\centering \includegraphics[height=5.5cm]{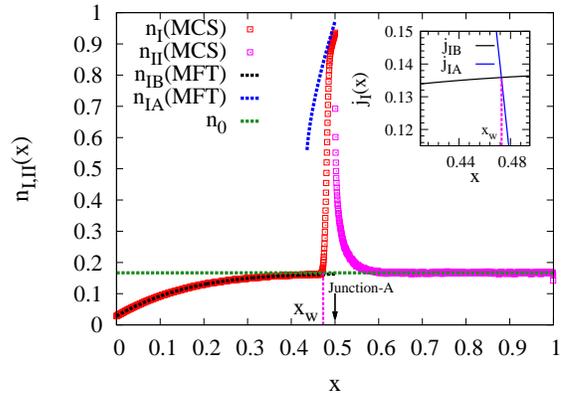}
\caption{(Colour online) Plots of the density profiles in SP  
($K=0.2,\,p=0.2,\,\Omega=4.5$, $l=1/2,\,N=2000$).  Mean-field solutions 
$n_{IB}(x)$ 
(black dotted lines), $n_{IA}(x)$ (blue dotted lines) and the corresponding 
MCS results $n_I(x)$ (red squares) are shown. MCS results for $n_{II}(x)$ are shown with magenta squares; $x_\alpha$ and $x_\beta$ 
are extracted from MCS data (see text). $n_0$ (green dotted lines) is the 
average density of the system. Inset: Mean-field values  of the 
currents $j_{IB}(x)$, $j_{IA}(x)$ and $j_0(x)$ are plotted; $x_\alpha$ and 
$x_\beta$ are extracted (see text), which match well with their corresponding 
MCS results.}
\label{K0.2_p0.2_om4.5}
\end{figure}
Given the physical expectation that $n_I(x)$ cannot decrease as $x$ rises, we 
identify two points $x_\alpha$ and $x_\beta$, at which $n_{IB}(x)$ and 
$n_{IA}(x)$, respectively, meet with the bulk solution $n_I(x)=n_0$. 
Defining $j_{IB}(x)$ and $j_{IA}(x)$ as the spatially varying 
currents corresponding to the 
densities $n_{IB}(x)$ and $n_{IA}(x)$, respectively: $j_{IB}(x)= 
n_{IB}(x)(1-n_{IB}(x))$, $j_{IA}(x)= 
n_{IA}(x)(1-n_{IA}(x))$ and $j_0(x)=n_0(1-n_0)$, 
$x_\alpha$ and $x_\beta$ may be obtained from $j_{IB}(x)=j_0$ and 
$j_{IA}(x)=j_0$. As in Ref.~\cite{LKTASEP_K=1}, three different scenarios are 
possible:

(i) {\em Continuous density phase} (CDP) corresponding to $x_\alpha<x_\beta$: 
$n_I(x)$ rises 
from $n_I(x=0)$ to reach $n_I(x)=n_0$ at $x_\alpha$, and then rise again from 
$x_\beta$ to reach $n_I(x)=n_I(x=l)$. Ignoring boundary layers, 
$n_{II}(x)=n_0$, i.e., GMC phase for CHII ensues. See Fig.~\ref{K0.5_p0.2_om4.5} for a 
representative plot with CDP for $n_I(x)$, where results 
from MFT and MCS studies are plotted together; good agreement between the MFT 
and MCS results are evident. Enumeration of $x_\alpha,\,x_\beta$ from MFT 
are shown in the inset. 
\par While with $K=1$, $n_I(x)$ takes a very simple form 
as a function of $x$~\cite{LKTASEP_K=1}, for $K\neq 1$ its functional form is 
more complex. Nonetheless, from the structures 
of Eqs.~\eqref{ch1_den}, \eqref{sol1n1} and \eqref{sol2n1}, we can make the 
following general observations.
(a) In general $n_{IA} > n_0=n_{II}(x)>n_{IB}$, (b) the slopes $\partial 
n_{IB,A}/\partial 
x\rightarrow 0$ as 
$n_{IB,A}(x)\rightarrow K/(1+K)$ at some points in the bulk, (c) with $K<1$, 
$\partial n_{IB}(x)/\partial x$ 
never diverges, where as $\partial n_{IA}(x)/\partial x$ diverges as 
$n_I(x)\rightarrow 1/2$. Thus broadly, the slope $ 
n_{IA}(x)$ should be {\em steeper} than that of $n_{IB}(x)$~\cite{comment1}. It is also clear that $n_{IA}(x)$ 
starts from $n_{IA}(x=x_\beta)=n_0 <1/2$ for $K<1$ to rise to 
(\ref{right_boundary}) that is larger than 1/2. Thus, $n_{IA}(x)$ is a 
combination of LD-HD phases. For $K=1$, $n_{IA}(x)$ is necessarily more than 
1/2, and hence fully HD, (d) lastly, $n_I(x)$  starts from LD  
(i.e., LD near junction B) and ends in HD (i.e., HD near junction A) always. 
These are consistent with our observations from  MCS results; see 
Fig.~\ref{K0.5_p0.2_om4.5}.

(ii) {\em Shock phase} (SP) corresponding to 
$x_\alpha>x_\beta$~\cite{LKTASEP_K=1}. There is no intervening flat portion in 
$n_I(x)$. Instead, $n_I(x)$ discontinuously changes from its value $n_{IB}$ to 
$n_{IA}$ at $x=x_w$, yielding a density shock or a localised domain wall (LDW) 
at $x=x_w$. The condition $j_{IB}(x_w)=j_{IA}(x_w)$ yields $x_w$. 
Density $n_{II}(x)$ remains at its GMC phase, i.e., $n_{II}(x)=n_0$, just like 
the CDP. See Fig.~\ref{ext_2ph} for a representative plot for 
SP of $n_I(x)$. Again, the agreement between MFT and MCS 
results is close. In the inset, enumeration of $x_w$ from MFT is shown.

(iii) The borderline case with $x_\alpha=x_\beta$: this corresponds to 
$j_{IB}(x)= j_{IA} (x) = j_0$ at $x=x_\alpha=x_\beta$. A representative profile 
of $n_I(x)$ with 
$x_\alpha=x_\beta$, i.e., at the boundary between SP and CDP, is shown in Fig. 
\ref{ext_bound}; inset shows a plot of numerical evaluation of 
$x_\alpha=x_\beta$ from the MFT. Unlike for $K=1$, 
when the 
density profiles at $x_\alpha=x_\beta$ are  linear~\cite{LKTASEP_K=1} (because 
the factor $1+\frac{2 b}{a}$
appearing in Eq. \eqref{eq:solution} becomes zero for $K=1$), for $K<1$, the 
same factor is not zero and hence we do not 
get a linear density profile at borderline case separating the CDP and 
SP regions. It may be noted that our above results hold good for 
all $l,\,0<l<1$; we have reported here the results only for $l=1/2$.
\par
\begin{figure}[t!]
\centering \includegraphics[height=5.5cm]{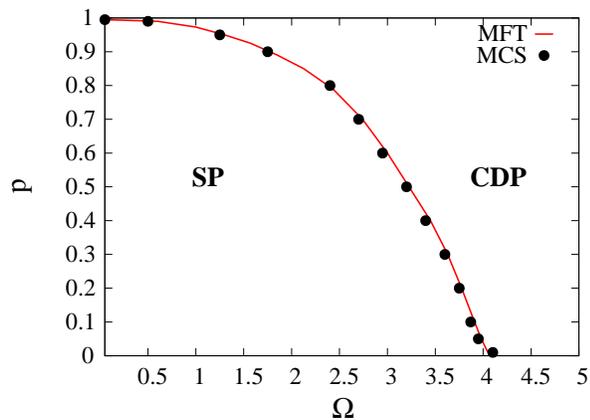}
\caption{(Colour online) Phase diagram in the $\Omega-p$ plane for extended 
defects, with 
$K=0.5$: mean field result (red continuous line) and MCS results (black 
circles) are shown, which agree well.  Here $l=1/2, N=1000$. }
\label{K0.5_ph_diag_ext}
\end{figure}

\begin{figure}[t!]
\centering \includegraphics[height=5.5cm]{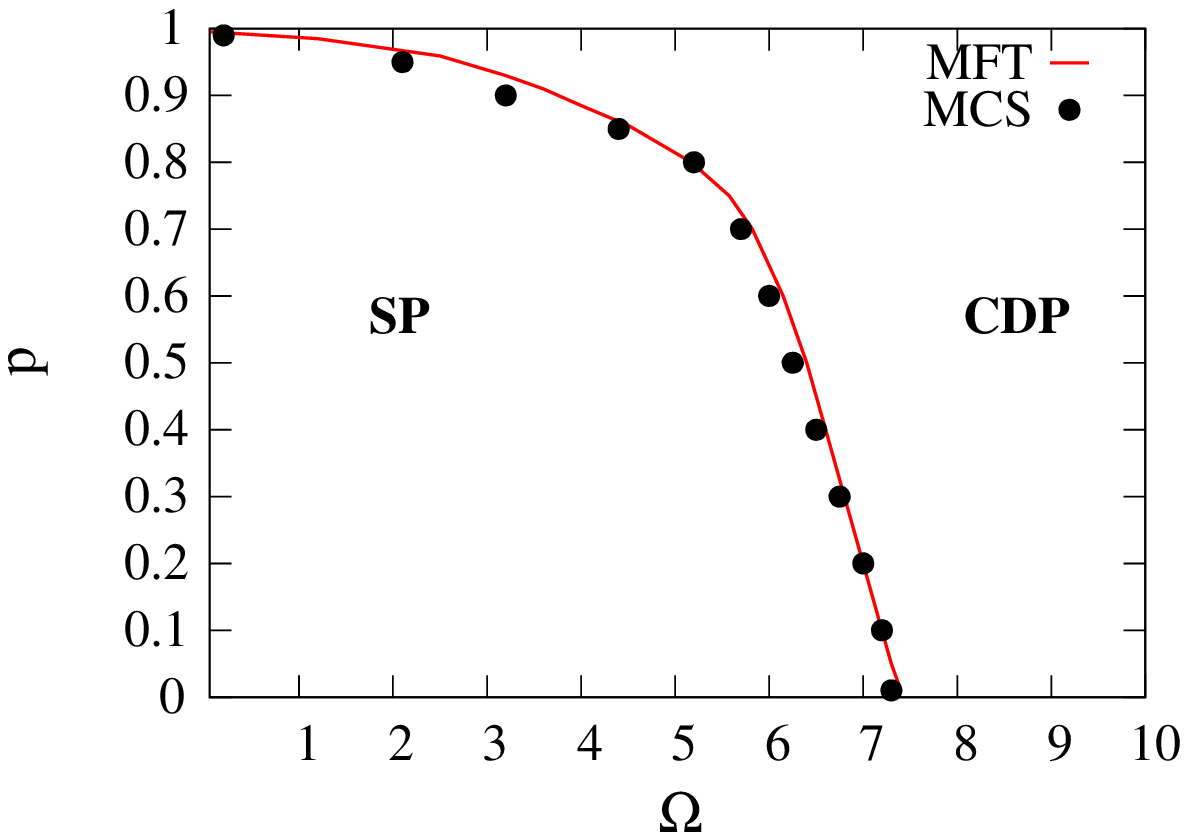}
\caption{(Colour online) Phase diagram in the $\Omega-p$ plane for extended 
defects, with 
$K=0.2$: mean field result (red continuous line) and MCS results (black 
circles) are shown, which agree well. Here $l=1/2, N=1000.$ }
\label{K0.2_ph_diag_ext}
\end{figure}

 In Figs.~\ref{K0.5_ph_diag_ext} and \ref{K0.2_ph_diag_ext}, the phase diagrams for $l=1/2$  in the 
$\Omega-p$ plane for $K=0.5$ and $K=0.2$, respectively, as obtained from our MCS and MFT 
approaches, are 
shown. Unsurprisingly, there are only two phases - CDP and SP in both of them. The phase 
boundary between CDP and SP are determined from the condition 
$x_\alpha=x_\beta$ (see above). 
Similar to Ref.~\cite{LKTASEP_K=1}, the MCS and MFT results mutually agree qualitatively as well 
as quantitatively.  The phase diagrams in Fig.~\ref{K0.5_ph_diag_ext} and Fig.~\ref{K0.2_ph_diag_ext}
are qualitatively similar to each other, and are also similar to the corresponding 
phase diagram for $K=1$ in Ref.~\cite{LKTASEP_K=1}. Still, Fig.~\ref{K0.5_ph_diag_ext} and 
Fig.~\ref{K0.2_ph_diag_ext}  do not 
match quantitatively, nor do they agree quantitatively with the corresponding 
phase diagram for $K=1$ in Ref.~\cite{LKTASEP_K=1}. 
\begin{table}
\begin{tabular}{|c|c|c|c|} 
 \hline
 K & 0.2 & 0.5 &  1.0 \\ 
 \hline
 $n_0$(MCS) &0.166672&0.333547&0.499187 \\ 
 \hline
 $n_0$(MFT) &0.166667&0.333333&0.5 \\
 \hline
\end{tabular}
\caption{Table comparing the average density value $n_0$ obtained from MCS and MFT for 
an extended defect. Here $\Omega=2.5$ and $p=0.5$, and $N=2000.$ }
\label{table:ext}
\end{table}
The region occupied 
by SP in Fig.~\ref{K0.2_ph_diag_ext} is noticeably 
larger than in Fig.~\ref{K0.5_ph_diag_ext}, which in turn is larger than the two-phase region for $K=1$; 
see Ref.~\cite{LKTASEP_K=1}. This trend may be explained in simple terms. 
From Eqs.~(\ref{left_boundary}) and (\ref{right_boundary}), it is clear that for small
$K$, the jump in the values of $n_I(x)$ across the extended defect is large. On the other hand,
$\partial n_I(x)/\partial x$ is clearly controlled by the factor $\Omega(1+K)$; see Eq.~(\ref{ch1_den}).
Now in order for the CDP to exist, $n_I(x)$ must rise (fall) sharply enough from its value $\alpha_1$ at $x=0$ 
($\alpha_2$ at $x=l$) to match with $n_0=K/(1+K)$ in the bulk. Since $\partial n_I(x)/\partial x$ is smaller for 
smaller $K$ for a given $\Omega$, a larger $\Omega$ is clearly needed with smaller $K$ to
ensure CDP. This explains why for $K=0.2$, the CDP requires a higher $\Omega$ 
than for $K=0.5$, which in turn has a higher-$\Omega$ threshold than that for three-phase coexistence for 
$K=1$~\cite{LKTASEP_K=1}. 
\par The changes in the phase diagrams given 
in Fig.~\ref{K0.5_ph_diag_ext} and Fig.~\ref{K0.2_ph_diag_ext} with variation 
in $K$ are also reflected in the corresponding plots of $n_I(x)$ versus $x$. 
For instance, compare Fig.~\ref{K0.5_p0.2_om4.5} 
($K=0.5$) with Fig.~\ref{K0.2_p0.2_om4.5} ($K=0.2$). Clearly, as $K$ is 
reduced a CDP density profile for $n_I(x)$ gives way to a SP density profile. 
This is consistent with the trends observed in the  above two phase diagrams.
\par In Table~\ref{table:ext}, we compare the mean density in the NESS of the 
whole system as obtained from MCS with the predictions from MFT for an 
extended defect. Clearly, good agreements are found. This validates our MFT arguments elucidated above.

\subsection{MF analysis and MCS results for a point defect}\label{1b}
\begin{figure}[t!]
\centering \includegraphics[height=5.5cm]{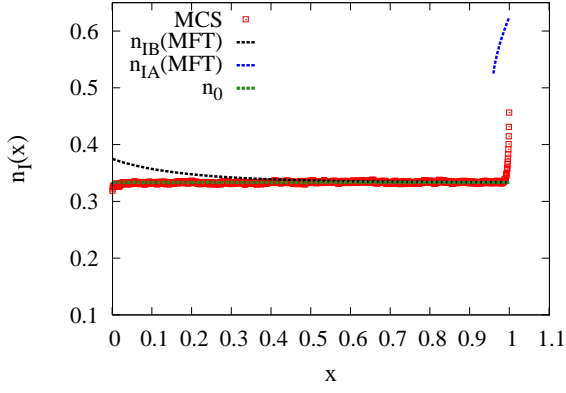}
\caption{(Colour online) Plots of the density profiles in the LD region for a 
point defect located at $x=1$
($K=0.5,\,p=0.6,\,\Omega=1.0,\, N=2000$).  Mean-field solutions $n_{IB}(x)$ 
(black dotted lines), $n_{IA}(x)$ (blue dotted lines) and the corresponding 
MCS results $n_I(x)$ (red squares) are shown. $n_0$ (green dotted lines) is the 
average density of the system. We have extended the $x$ axis up to $x=1.1$ 
to resolve the peak at $x=1$ (the location of the point defect) properly.}
\label{pd_LD}
\end{figure}

For a point defect, junctions A and B merge. As a 
result, the constraint from the 
constant bulk current in 
CHII on $j_I(x)=n_I(x)(1-n_I(x))$ for an 
extended defect 
does not exist for a point defect. In fact, CHII effectively 
ceases to
exist, and Eq.~\eqref{ch2_den} no longer holds. The steady state 
density $n_I(x)$ of CHI 
now spans the whole system, and satisfies Eq. \eqref{ch1_den}, 
\beq
\left(2 n_I(x)-1\right) \frac{\partial n_{I}(x)}{\partial x} - \Omega (1+ K)  
\left(n_{I}(x)-\frac{K}{1+K}\right) = 0.
\label{point_den}
\eeq
We now argue that (\ref{point_den}) allows for a 
spatially uniform steady state 
density in CHI, with a localised peak at the location of the point defect. 
Evidently, (\ref{point_den}) allows for a uniform solution 
$n_I(x)=K/(1+K)<1/2$ ($K<1$), in addition to the space-dependent solutions, 
akin to the solutions of $n_I(x)$ in the extended defect case, in the NESS of 
the model. As argued in Refs.~\cite{erwin2,niladri}, for a sufficiently low average 
density, the effect of the point defect is confined to creating a localised peak (or a 
dip) that has a vanishing width in the thermodynamic limit, in an otherwise 
homogeneous density profiles. In other words, the bulk density should be 
uniform for sufficiently low average density. For a ring geometry, this implies 
$n_I(x)=K/(1+K)$, with $n_I(x)=n_0+h$ having a localised peak of height 
$h=n_0(1-p)/p$ at the 
location of the point defect~\cite{erwin2,niladri}. This is consistent with our 
discussions above. See Fig.~\ref{pd_LD} for a representative plot of $n_I(x)$ 
as a function of $x$ in the LD phase. As $n_0$ rises (i.e., $K$ 
rises), or the defect strength rises (i.e., $p$ decreases), eventually this 
picture breaks down and the point defect then leads to macroscopically nonuniform 
density profiles. Following the logic outlined in 
Refs.~\cite{erwin2,niladri}, we find the 
threshold of inhomogeneous phases to be
\begin{figure}[t!]
\centering \includegraphics[height=5.5cm]{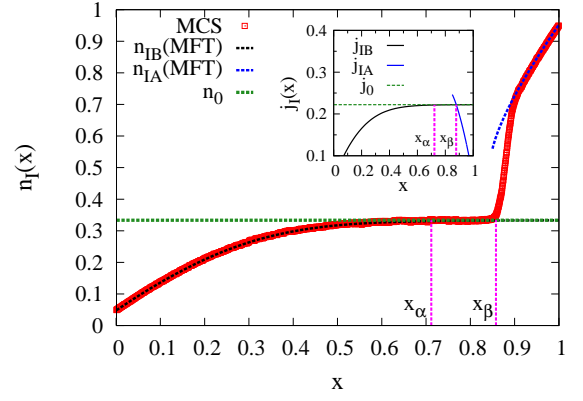}
\caption{(Colour online) Plots of the density profiles in the CDP 
($K=0.5,~p=0.05,~\Omega=2.0,\, N=2000$).  Mean-field solutions $n_{IB}(x)$ 
(black dotted lines), $n_{IA}(x)$ (blue dotted lines) and the corresponding 
MCS results $n_I(x)$ (red squares) are shown. MCS results for $n_{II}(x)$ are shown with magenta squares; $x_w$ is 
extracted from MCS data (see text). $n_0$ (green dotted lines) is the 
average density of the system. Inset: Mean-field values  of the 
currents $j_{IB}(x)$, $j_{IA}(x)$  are plotted; $x_w$ has been extracted (see 
text). Qualitative agreement with their corresponding 
MCS results are evident.}
\label{pdK0.5_CDP}
\end{figure}
\begin{equation}
\frac{K}{1+K}=\frac{p}{1+p}\implies K=p.\label{pointcond}
\end{equation}
Therefore as $K$ exceeds $p$, spatially nonuniform density profiles are to be 
formed in the NESS. 
\par For $K>p$,  the spatially varying solutions of $n_I(x)$ may be analysed as 
before. Without any loss 
of generality, we assume that the site
with the defect is located at $x=1$. Assuming that the particles hop 
anti-clockwise as before, one expects 
that $n(1-\epsilon)>n(\epsilon)$,
where $\epsilon \to 0$. Let $\rho_L$ and $\rho_{R}$ be the densities at the 
left and right of $x=1$. Now assume phase coexistence for $n_I(x)$ in the NESS, 
i.e., no boundary layers at $x=1$. Applying the current conservation at $x=1$, 
we get
\beq
\rho_L (1-\rho_L) = p \rho_{R}(1-\rho_L)= \rho_{R}(1-\rho_{R}).
\label{defect_current_cons}
\eeq
This gives, $\rho_L=\frac{p}{1+p}$ and $\rho_{R}=\frac{1}{1+p}$. The CHI 
density is then 
obtained by using Eq.~(\ref{ch1_den}) along with the boundary conditions
\beq
n_I(0)=\frac{p}{1+p}
~~\text{and} ~~n_I(1)=\frac{1}{1+p}.
\label{pd_bound_cond}
\eeq
As for the extended defect case, there are three different solutions: two 
spatially varying depending upon the boundary conditions (\ref{pd_bound_cond}) 
and a third bulk solution $n_I(x)=K/(1+K)=n_0$, that is independent of $x$. The 
spatially varying solutions are
{\small
\beq
2 \left(n_{IB}(x)-\frac{p}{1+p}\right)- \left( 1+ \frac{2 b}{a}\right) \ln 
\left|\frac{a n_{IB}(x)+ b}{\frac{a p}{1+p}+ b}\right|= ax, \label{pd_sol1n1}
\eeq
and 
\beq
2 \left(n_{IA}(x)-\frac{1}{1+p}\right)- \left( 1+ \frac{2 b}{a}\right) \ln 
\left|\frac{a n_{IA}(x)+ b}{\frac{a}{1+p}+ b}\right|= a(x-1).\label{pd_sol2n1}
\eeq}
Again as in CDP with extended defect, we can define $x_\alpha$ and $x_\beta$ 
from the conditions $n_{IB}(x_\alpha)=n_0$ and $n_{IA}(x_\beta)=n_0$. 
Equivalently, defining currents $j_{IA}(x)=n_{IA}(x)[1-n_{IA}(x)]$, 
$j_{IB}(x)=n_{IB}(x)[1-n_{IB}(x)]$ and $j_0(x)=n_0(1-n_0)$, $x_\alpha$ and 
$x_\beta$ are obtained from $j_{IB}(x_\alpha)=j_0$ and $j_{IA}(x_\beta)=j_0$ respectively.
Similar to the extended defect case (see also Ref.~\cite{LKTASEP_K=1}), three 
distinct cases are possible: 

(i) CDP for $x_\alpha<x_\beta$: $n_I(x)$ is qualitatively 
similar to its analogue for the extended defect case. A representative plot of 
$n_I(x)$ in its CDP as a function of $x$ is shown in 
Fig.~\ref{pdK0.5_CDP}.

\begin{figure}[t!]
\centering \includegraphics[height=5.5cm]{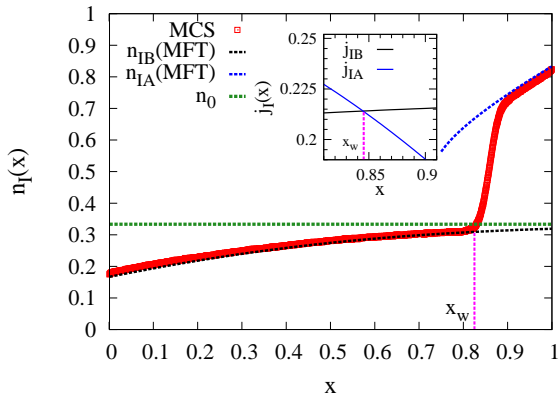}
\caption{(Colour online) Plots of the density profiles in the SP phase
($K=0.5,~p=0.2,~\Omega=0.75,\, N=2000$).  Mean-field solutions $n_{IB}(x)$ 
(black dotted lines), $n_{IA}(x)$ (blue dotted lines) and the corresponding 
MCS results $n_I(x)$ (red squares) are shown. MCS results for $n_{II}(x)$ are shown with magenta squares; $x_w$ is 
extracted from MCS data (see text). $n_0$ (green dotted lines) is the 
average density of the system. Inset: Mean-field values  of the 
currents $j_{IB}(x)$, $j_{IA}(x)$  are plotted; $x_w$ has been extracted (see 
text). Qualitative agreement with their corresponding 
MCS results are evident.}
\label{pd_2ph}
\end{figure}

(ii) SP with $x_\alpha>x_\beta$ yielding an LDW at $x=x_w$. A 
representative plot of 
$n_I(x)$ in its SP as a function of 
$x$ with an LDW at $x=x_w$ is shown in 
Fig.~\ref{pd_2ph}.

\begin{figure}[t!]
\centering \includegraphics[height=5.5cm]{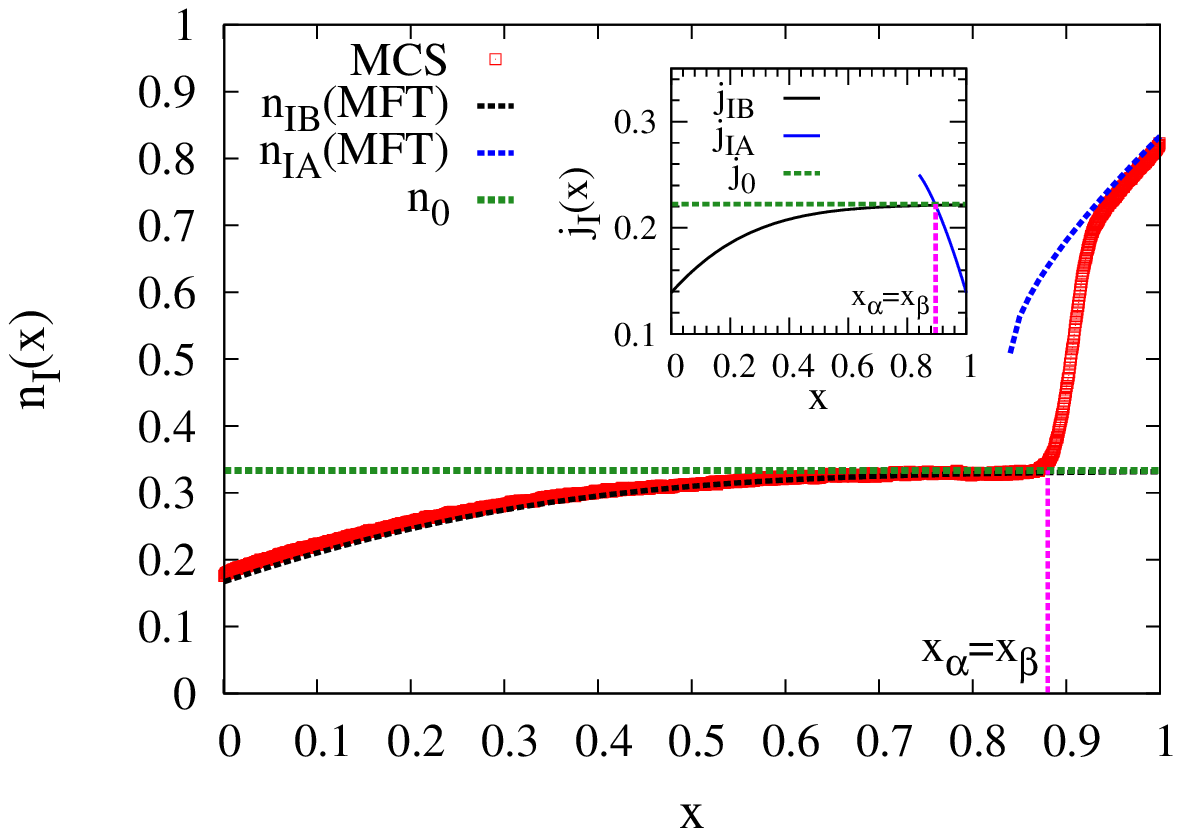}
\caption{(Colour online) Plots of the density profiles in the borderline case 
for a point defect located at $x=1$
($K=0.5,\,p=0.2,\,\Omega=1.25,\, N=2000$).  Mean-field solutions $n_{IB}(x)$ 
(black dotted lines), $n_{IA}(x)$ (blue dotted lines) and the corresponding 
MCS results $n_I(x)$ (red squares) are shown. MCS results for $n_{II}(x)$ are shown with 
magenta squares; $x_\alpha$ and $x_\beta$ 
are extracted from MCS data (see text). $n_0$ (green dotted lines) is the 
average density of the system. Inset: Mean-field values  of the 
currents $j_{IB}(x)$, $j_{IA}(x)$ and $j_0(x)$ are plotted; $x_\alpha$ and 
$x_\beta$ are extracted (see text), which match well with their corresponding 
MCS results.}
\label{pd_bound23}
\end{figure}

(iii) The borderline case between CDP and SP given by 
$x_\alpha=x_\beta$; see Fig.~\ref{pd_bound23}.
\par
\begin{figure}[t!]
\centering \includegraphics[height=5.5cm]{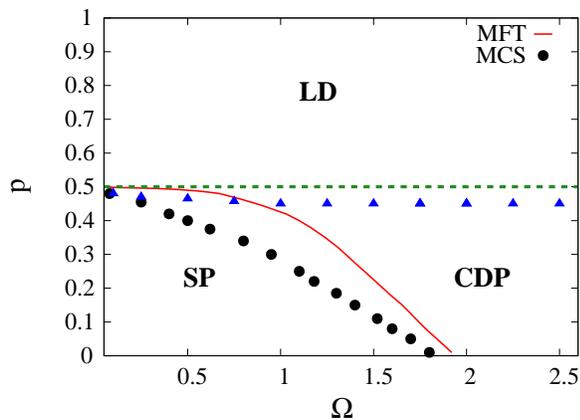}
\caption{ Phase diagram in the $\Omega-p$ plane for a point defect located 
at $x=1$, with $K=0.5.$: Mean-field results (red solid line) and the MCS results 
are shown. The green-dotted line, $K=p$ separating the LD-phase
from other phase regions is obtained from MFT and satisfy the equation $p=K$. 
The corresponding MCS data points are represented by the blue-triangles. Here $N=1000$.}
\label{K0.5_ph_diag_pd}
\end{figure}

\begin{figure}[t!]
\centering \includegraphics[height=5.5cm]{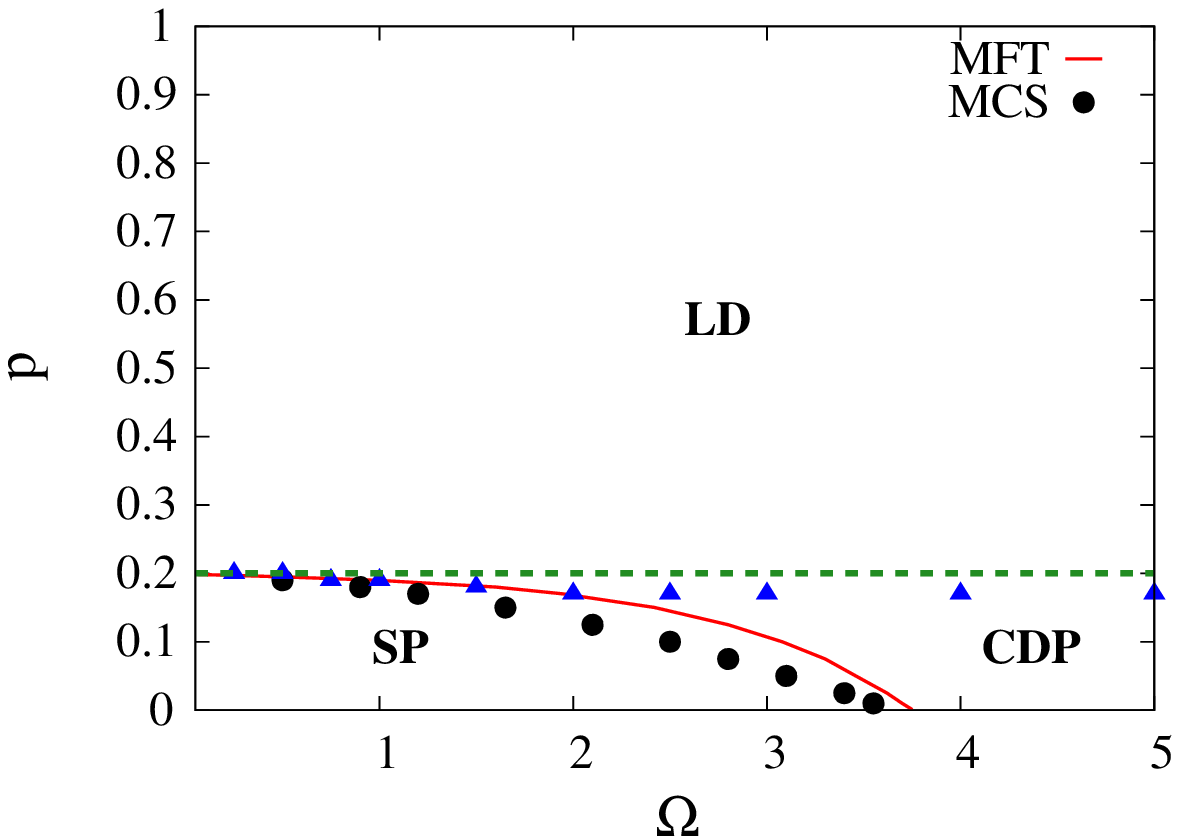}
\caption{ Phase diagram in the $\Omega-p$ plane for a point defect located 
at $x=1$, with $K=0.2.$: Mean-field results (red solid line) and the MCS results 
are shown. The green-dotted line, $K=p$ separating the LD-phase
from other phase regions is obtained from MFT and satisfy the equation $p=K$. 
The corresponding MCS data points are represented by the blue-triangles. Here $N=1000$.}
\label{K0.2_ph_diag_pd}
\end{figure}

 To find out the phase boundary between the CDP and SP 
regions, we proceed in the same way as in case of extended defects, that is we 
find out $p$ and $\Omega$ values for which $x_\alpha=x_\beta$. For $K<p$, 
the system is in a spatially homogeneous LD phase. The phase diagram for the case of point defect showing the three 
distinct phase regions is shown in Figs.~\ref{K0.5_ph_diag_pd} and \ref{K0.2_ph_diag_pd}. 
\begin{figure}[t!]
\centering \includegraphics[height=5.5cm]{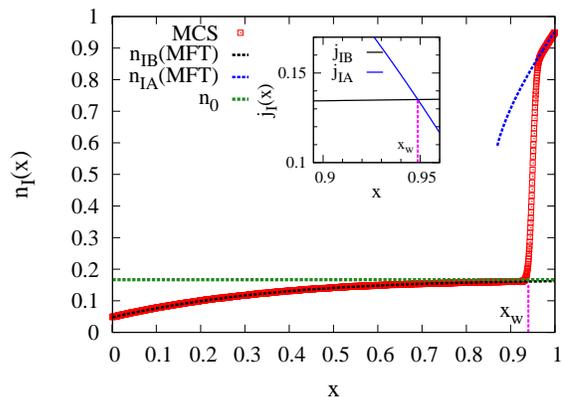}
\caption{(Colour online) Plots of the density profiles in the SP for a point 
defect located at $x=1$
($K=0.2,\,p=0.05,\,\Omega=2.0,\, N=2000$).  Mean-field solutions $n_{IB}(x)$ 
(black dotted lines), $n_{IA}(x)$ (blue dotted lines) and the corresponding 
MCS results $n_I(x)$ (red squares) are shown. MCS results for $n_{II}(x)$ are shown with 
magenta squares; $x_\alpha$ and $x_\beta$ 
are extracted from MCS data (see text). $n_0$ (green dotted lines) is the 
average density of the system. Inset: Mean-field values  of the 
currents $j_{IB}(x)$, $j_{IA}(x)$ and $j_0(x)$ are plotted; $x_\alpha$ and 
$x_\beta$ are extracted (see text), which match well with their corresponding 
MCS results.}
\label{pd_K0.2_p0.05_om2.0}
\end{figure}
\par Notice that the extent of the LD phase in Fig.~\ref{K0.2_ph_diag_pd} is 
distinctly larger than in Fig.~\ref{K0.5_ph_diag_pd}. This is consistent with 
the fact that the boundary between LD and the spatially inhomogeneous phases 
(SP or CDP) in MFT is given by $p=K$ (in agreement with MCS results), that 
clearly yields a larger LD phase in the $\Omega-p$ plane for a lower $K$. Next, 
consider the relative preponderance of the SP over the CDP in 
Fig.~\ref{K0.2_ph_diag_pd} in comparison with Fig.~\ref{K0.5_ph_diag_pd}.
Unlike the case with an extended defect, the jump in an inhomogeneous $n_I(x)$ 
at the point defect depends only on $p$, and not on $K$, regardless of SP or 
CDP. Nonetheless, as discussed above, the slope $\partial n_I(x)/\partial x$ in 
NESS is still controlled by $\Omega(1+K)$. Hence, for reasons similar to those 
for an extended defect, in a $\Omega-p$ phase diagram, the CDP starts for a 
higher $\Omega$ with a lower $K$. For $K=1$, there is no LD phase even for a 
point defect. For a putative LD phase one must have $p>K$; for $K=1$ this 
condition rules out an LD phase. 
\par The quantitative dissimilarities between the 
phase diagram in Ref.~\cite{LKTASEP_K=1} for a point defect and the 
inhomogeneous parts of those in Figs.~\ref{K0.5_ph_diag_pd} and 
\ref{K0.2_ph_diag_pd} may be explained following the logic outlined in the 
discussions on the phase diagrams for an extended defect above. These differences can lead to qualitative differences in the density profiles 
in the NESS. For instance, compare Fig.~\ref{pdK0.5_CDP} ($K=0.5$) and Fig.~\ref{pd_K0.2_p0.05_om2.0} ($K=0.2$) to 
observe that the CDP density profile for $K=0.5$ has become a SP density profile for $K=0.2$. This trend
is similar to what one finds for extended defects.
\par In Table~\ref{table:pd}, we compare the values of the mean densities obtained 
from MCS with the corresponding MFT results for different values of $K$. 
Again, similar to the extended defect case, there is a good quantitative 
agreement between the two, lending credence to our MFT results.

\section{Summary and outlook}\label{summ}
In this work, we have studied an asymmetric exclusion  process in an 
inhomogeneous ring with particle nonconserving LK dynamics. The attachment/detachment 
rates of LK are generally assumed to be unequal. We have considered both 
extended and point defects. 
\begin{table}
\begin{tabular}{|c|c|c|c|} 
 \hline
 K & 0.2 & 0.5 &  1.0 \\ 
 \hline
 $n_0$(MCS) &0.166915&0.333366&0.499587 \\ 
 \hline
 $n_0$(MFT) &0.166667&0.333333&0.5 \\
 \hline
\end{tabular}
\caption{Table comparing the average density value $n_0$ obtained from MCS and 
MFT for a point defect. Here $\Omega=1.0$, $p=0.5$ and $N=2000$.}
\label{table:pd}
\end{table}
The MFT analysis is done by assuming the system 
to be a combination of two TASEP channels CHI (of hopping rate 1) and CHII 
(of hopping rate $p<1$) of unequal hopping rates, which are joined together at 
both the ends. For a point defect, CHII shrinks to a point. Our MFT analysis, backed up by 
extensive MCS studies,  clearly reveals that with an extended defect, 
$n_{II}=K/(1+K)$ in the bulk of CHII, a constant that is independent of $p(<1)$ 
(i.e., independent of 
the boundaries or junctions at A and B). In analogy with the MC phase of an 
open TASEP, we call this the GMC phase. In contrast, CHI is found in either 
CDP or SP. Unsurprisingly, CDP is 
favoured for larger $\Omega$, where as SP prevails for smaller $\Omega$ for  
fixed $p, K$, as in Ref.~\cite{LKTASEP_K=1}. For a point defect, 
for which CHII shrinks to a point, for $K<p$, $n_I(x)$ is found in a uniform 
phase, unlike an extended defect. The variation in the phase diagrams with 
$K$ are explained in simple physical terms. As in Ref.~\cite{LKTASEP_K=1}, the 
quantitative agreement between the MCS and MFT results for an extended defect 
is very high, as is evident in the corresponding phase diagram; see 
Fig.~\ref{K0.5_ph_diag_ext} and Fig.~\ref{K0.2_ph_diag_ext}. In contrast, the 
agreement for the point defect case is weaker, 
in particular for small $\Omega$. The physical reason behind these 
discrepancies are expected to be the same as that elaborated in 
Ref.~\cite{LKTASEP_K=1}.
\par Our results
clearly bring out the relevance of the ring or closed
geometry of the system in the presence of LK with unequal attachment and 
detachment rates. The results here as well as those in 
Ref.~\cite{LKTASEP_K=1} clearly establish how the ring geometry (or the lack 
of independent entry and exit events) restricts the possible phases in these 
models, in comparison with the results on the corresponding open system; see 
Refs.~\cite{erwin-lk,erwin2}. It would be interesting to numerically study the 
crossover between the extended and point defect cases by varying $N$, while 
keeping the length of the slower segment unchanged. As an alternative to our 
simple MFT, it should interesting to extend the boundary layer formalism 
developed in Refs.~\cite{somen} for the present problem. The simplicity of our 
model limits direct applications of our results to practical or experimental 
situations. Nonetheless, our above results in the context of vehicular traffic along a 
circular track, or railway movements in series along a closed railway loop line 
with possibilities of new carriages joining or existing carriages going off the 
loop track in the 
presence of constrictions (regions of slow passages due to, e.g., accidents or damages 
in the tracks), or ribosome translocations
along mRNA loops with defects and random attachment/detachment, clearly 
demonstrate that for extended defects, the steady state densities are in 
general inhomogeneous, where as for a point defect, globally uniform densities 
are possible for a sufficiently low average density. We hope
experiments on ribosomes using ribosome profiling techniques~\cite{exp1} and 
numerical simulations of more detailed
traffic models should qualitatively validate our results.

\section{Acknowledgement} \label{ack}
SM acknowledges the financial support from the Council of 
Scientific and Industrial Research, India [Grant No. 09/489(0096)/2013-EMR-I].
TB and AB gratefully acknowledge partial financial support from Alexander von 
Humboldt Stiftung, Germany under the Research Group 
Linkage Programme (2016).


\begin{thebibliography}{99}
\bibitem{equi-disord} J. Ziman, {\em Models of Disorder: the Theoretical 
Physics of Homogeneously Disordered Systems} (Cambridge University Press, Cambridge, MA, 
USA, 1979); R. Stinchcombe, {\em Dilute magnetism}, vol. 7 of Phase Tran-
sition and Critical Phenomena (Academic Press, New
York, NY, USA, 1983).
%%%%%%%%%%%%%%%%%%%%%%%%%%%%%%%%%%%%%%%%%%%%%%%%%%%%%%%%%%%%%%%%%%%%%%%%%%%%%
\bibitem{noneq-quench} R. Stinchcombe, J. Phys.: Condens. Matter {\bf14}, 1473 (2002).
%%%%%%%%%%%%%%%%%%%%%%%%%%%%%%%%%%%%%%%%%%%%%%%%%%%%%%%%%%%%%%%%%%%%%%%%%%%%%%
\bibitem{review} D. Chowdhury, L. Santen, and A. Schadschneider, Phys.
Rep. {\bf 329}, 199 (2000); D. Helbing, Rev. Mod. Phys. {\bf 73},
1067 (2001); T. Chou, K. Mallick and R. K. P. Zia, {\em Rep. Prog. Phys.} {\bf 74},
116601 (2011).
%%%%%%%%%%%%%%%%%%%%%%%%%%%%%%%%%%%%%%%%%%%%%%%%%%%%%%%%%%%%%%%%%%%%%%%%%%%%%%%%
\bibitem{nuclearpore} I. Kosztin and K. Schulten, Phys. Rev. Lett. {\bf 93},
238102 (2004).
%%%%%%%%%%%%%%%%%%%%%%%%%%%%%%%%%%%%%%%%%%%%%%%%%%%%%%%%%%%%%%%%%%%%%%%%%%%%
\bibitem{molmot} C. T. MacDonald, J. H. Gibbs and A. C. Pipkin, Biopolymers
{\bf 6}, 1 (1968); R. Lipowsky, S. Klump and T. M. Nieuwenhuizen, Phys. Rev. Lett. {\bf 87}, 108101 (2001).
%%%%%%%%%%%%%%%%%%%%%%%%%%%%%%%%%%%%%%%%%%%%%%%%%%%%%%%%%%%%%%%%%%%%%%%%%%%%
\bibitem{zeo}J. K\"arger and D. Ruthven,
Diffusion in Zeolites and other microporous solids (Wiley, New York, 1992)
%%%%%%%%%%%%%%%%%%%%%%%%%%%%%%%%%%%%%%%%%%%%%%%%%%%%%%%%%%%%%%%%%%%%%%%%%%%%
\bibitem{albertbook} B. Alberts {\em et al}, {\em Molecular Biology of the 
Cell}, Garland Science, New York (2002).
\bibitem{erwin-lk} A. Parmeggiani, T. Franosch and E. Frey, Phys. Rev. E {\bf 70}, 046101 (2004).
%%%%%%%%%%%%%%%%%%%%%%%%%%%%%%%%%%%%%%%%%%%%%%%%%%%%%%%%%%%%%%%%%%%%%%%%%%%%%%
\bibitem{tasep-def}J. J. Dong, B. Schmittmann, and R. K. P. Zia, J. Stat. Phys.
{\bf 128}, 21 (2007); P. Greulich and A. Schadschneider, Physica A {\bf 387}, 
1972 (2008); R. K. P. Zia, J. J. Dong, and B. Schmittmann, J. Stat.
Phys. {\bf 144}, 405 (2011); J. S. Nossan, J. Phys. A: Math. Theor. {\bf 46}, 
315001 (2013).
%%%%%%%%%%%%%%%%%%%%%%%%%%%%%%%%%%%%%%%%%%%%%%%%%%%%%%%%%%%%%%%%%%%%%%%%%%%%%%%
%%%%%%%%%%%%%%%%%%%%%%%%%%%%%%%%%%%%%%%%%%%%%%%%%%%%%%%%%%%%%%%%%%%%%%%%%%%%%%%
\bibitem{erwin2} P. Pierobon, M. Mobilia, R. Kouyos, and E. Frey, 
Phys. Rev. E {\bf 74}, 031906 (2006).
%%%%%%%%%%%%%%%%%%%%%%%%%%%%%%%%%%%%%%%%%%%%%%%%%%%%%%%%%%%%%%%%%%%%%%%%%%%%%%
\bibitem{Mustansir1} G. Tripathy and M. Barma, Phys. Rev. E {\bf 58}, 1911 (1998)
%%%%%%%%%%%%%%%%%%%%%%%%%%%%%%%%%%%%%%%%%%%%%%%%%%%%%%%%%%%%%%%%%
\bibitem{niladri} N. Sarkar and A. Basu,  Phys. Rev. E {\bf 90}, 022109 (2014).
%%%%%%%%%%%%%%%%%%%%%%%%%%%%%%%%%%%%%%%%%%%%%%%%%%%%%%%%%%%%%%%%%%
\bibitem{tirtha1} T. Banerjee, N. Sarkar and A. Basu, J.
Stat. Mech.: Theory and Experiment, P01024 (2015).
%%%%%%%%%%%%%%%%%%%%%%%%%%%%%%%%%%%%%%%%%%%%%%%%%%%%%%%%%%%%%%%%%
\bibitem{LKTASEP_K=1}  T. Banerjee, A. K. Chandra, and A. Basu, Phys. Rev. E 
{\bf 92}, 022121 (2015).
%%%%%%%%%%%%%%%%%%%%%%%%%%%%%%%%%%%%%%%%%
\bibitem{comment1} Notice that this condition cannot be used for $K=1$ as in 
Ref.~\cite{LKTASEP_K=1}, where the factor $2n_I(x)-1$ vanishes as $n_I(x) 
\rightarrow 1/2$, making the slope $\partial n_I(x)/\partial x$ indeterminate. 
This suggests a discontinuity in the slope, unlike here, where the slope 
changes continuously. This is borne out by our MCS studies as well.
%%%%%%%%%%%%%%%%%%%%%%%%%%%%%%%%%%%%%%%
\bibitem{gmc-k} For $K>1$, $n_{II}(x)=K/(1+K)>1/2$ always.
%%%%%%%%%%%%%%%%%%%%%%%%%%%%%%%%%%%%%%%%%%%%%%%%%%%%%%%%%
\bibitem{somen} S. Mukerjee and S. M. Bhattacharjee, J. Phys. A: Math.
Gen. {\bf 38}, L285 (2005); S. Mukherji and V. Mishra, Phys.
Rev. E {\bf 74}, 011116 (2006); S. M. Bhattacharjee, J. Phys.
A: Math. Theor. {\bf 40}, 1703 (2007); S. Mukherji, Phys.
Rev. E {\bf 79}, 041140 (2009); S. Mukherji, Phys. Rev. E
{\bf 83}, 031129 (2011); A. K. Gupta and I. Dhiman, Phys. Rev. E {\bf 89}, 022131 (2014).
%%%%%%%%%%%%%%%%%%%%%%%%%%%%%%%%%%%%%%%%%%%%%%%%%%%%%%%%%%%%%%%%%%
\bibitem{exp1} Y. Arava {\em et al}, Nucl. Acids Res. 33, 2421 (2005); N.
T. Ingolia {\em et al}, {\em Science} {\bf 324}, 218 (2009); H. Guo {\em et al},
Nature {\bf 466}, 835 (2010).
%%%%%%%%%%%%%%%%%%%%%%%%%%%%%%%%%%%%%%%%%%%%%%%%%%%%%%%%%%%%%%%%
\end{thebibliography}
\end{document}